\documentclass[apj]{emulateapj}
\usepackage{mathptmx}
\usepackage{apjfonts}

\newcommand  \kms      {\ifmmode {\rm km\,s}^{-1} \else km\,s$^{-1}$\fi}
\newcommand  \ergs     {\ifmmode {\rm ergs\,s}^{-1} \else ergs s$^{-1}$\fi}
\newcommand  \ergcms   {\ifmmode {\rm ergs\,cm}^{-2}\,{\rm s}^{-1}
                        \else ergs\,cm$^{-2}$\,s$^{-1}$\fi}
\newcommand  \ergcmsA  {\ifmmode{\rm ergs\,cm}^{-2}\,{\rm s}^{-1}\,{\rm\AA}^{-1}
                        \else ergs\,cm$^{-2}$\,s$^{-1}$\,\AA$^{-1}$\fi}

\journalinfo{The Astrophysical Journal, 
6??: ??1--?11, 2006 Month dd, Preprint on astro-ph}
\slugcomment{Received 2006 September 7; accepted 2006 December 19}

\shorttitle{REVERBERATION MAPPING OF HIGH-LUMINOSITY QUASARS}

\shortauthors{KASPI ET Al.}

\begin{document}

\title{Reverberation Mapping of High-Luminosity Quasars: First Results}

\author{
Shai~Kaspi,\altaffilmark{1,2} 
W. N. Brandt,\altaffilmark{3}
Dan~Maoz,\altaffilmark{1} 
Hagai~Netzer,\altaffilmark{1} \\ 
Donald P. Schneider,\altaffilmark{3}
and
Ohad Shemmer\altaffilmark{3}
}

\altaffiltext{1}{School of Physics and Astronomy, Raymond and Beverly
Sackler Faculty of Exact Sciences, Tel-Aviv University, Tel-Aviv 69978,
Israel; shai@wise.tau.ac.il.}
\altaffiltext{2}{Physics Department, Technion, Haifa 32000, Israel.}
\altaffiltext{3}{Department of Astronomy and Astrophysics, 525 Davey
Laboratory, Pennsylvania State University, University Park, PA 16802.}

\begin{abstract}
Reverberation mapping of nearby active galactic nuclei has led
to estimates of broad-line-region (BLR) sizes and central-object
masses for some 37 objects to date. However, successful reverberation
mapping has yet to be performed for quasars of either high luminosity
(above $L_{\rm opt}\sim 10^{46}~{\rm erg~s}^{-1}$) or high redshift
($z\gtrsim$\,0.3). Over the past six years, we have carried out, at
the Hobby-Eberly Telescope, rest-frame-ultraviolet spectrophotometric
monitoring of a sample of six quasars at redshifts $z=2.2-3.2$,
with luminosities of $L_{\rm opt}\sim 10^{46.4}$--$10^{47.6}~{\rm
erg~s}^{-1}$, an order of magnitude greater than those of previously
mapped quasars. The six quasars, together with an additional five
having similar redshift and luminosity properties, were monitored
photometrically at the Wise Observatory during the past decade.
All 11 quasars monitored show significant continuum variations of order
10--70\%. This is about a factor of two smaller variability than for
lower luminosity quasars monitored over the same rest-frame period.
In the six objects which have been spectrophotometrically monitored,
significant variability is detected in the \ion{C}{4}$\lambda1550$
broad emission line.  In several cases the variations track the
continuum variations in the same quasar, with amplitudes comparable to,
or even greater than, those of the corresponding continua. In contrast,
no significant Ly$\alpha$ variability is detected in any of the four
objects in which it was observed. Thus, UV lines may have different
variability trends in high-luminosity and low-luminosity AGNs.  For one
quasar, S5\,0836+71 at $z=2.172$, we measure a tentative delay of 595
days between \ion{C}{4} and UV-continuum variations, corresponding
to a rest-frame \ delay of 188 days and a central \ black-hole mass of
$2.6\times10^9 M_{\odot}$.

\end{abstract}


\section{Introduction}

Reverberation mapping has been used in the past two decades to estimate
the sizes of the broad-line-emitting regions (BLRs) in several dozens
of active galactic nuclei (AGNs), and thus to infer the masses of the
black holes at their centers (e.g., Kaspi et al. 2000). The technique
utilizes the light-travel-time delayed flux response of the BLR to
changes in the continuum flux (for reviews, see Peterson 1993; Netzer
\& Peterson 1997; Peterson 2006 and references therein). Recently, a
compilation of all available reverberation-mapping data, for 37 AGNs,
was analyzed in a uniform and self-consistent manner. The relationship
between luminosity, $L$, and the BLR size, $R_{\rm BLR}$, in several
luminosity bands was studied by Kaspi et al. (2005), and the black-hole
mass-luminosity relation, $M_{\rm BH}-L$, based on these data was
re-derived by Peterson et al. (2004). Current reverberation studies
cover four orders of magnitude in AGN luminosity in a well-sampled
manner, from $\lambda L_\lambda (5100{\rm\AA})\sim 10^{42}$ \ergs\
to $10^{46}$ \ergs . In this range, $R_{\rm BLR} \propto L^\alpha$,
with $\alpha=0.67\pm0.05$ for the optical continuum, confirming
the relations found by Kaspi et al. (2000; c.f., Bentz et al. 2006
find $\alpha=0.518\pm0.039$) and $M_{\rm BH}\propto L^\beta$, with
$\beta=0.79 \pm 0.09$. In a recent study Peterson et al. (2005)
measured the BLR size of the \ion{C}{4}$\lambda$1550 emission line in
the least-luminous AGN, NGC\,4395, which has an optical luminosity
of $\lambda L_\lambda$(5100\,\AA )=5.9$\times 10^{39}$ \ergs. They
find the \ion{C}{4} BLR size to be $1\pm 0.3$ light hr, consistent
with the size expected from extrapolating the $R_{\rm BLR}$--$L$
relation to lower luminosities.

\begin{deluxetable*}{lrccccccccrc}
\tablecolumns{12}
\tabletypesize{\scriptsize}
\tablewidth{430pt}
\tablecaption{Object Characteristics
\label{sample}}
\tablehead{
\colhead{Object} &
\colhead{RA (2000)} &
\colhead{Dec (2000)} &
\colhead{$m_V$} &
\colhead{Redshift} &
\colhead{$N_{\rm phot}$} &
\colhead{$N_{\rm spec}$} &
\colhead{$\lambda L_{\lambda}$(5100\AA )} &
\colhead{$R_{\rm comp}$} &
\colhead{$PA_{\rm comp}$} &
\colhead{$R$} &
\colhead{Exp.} \\
\colhead{(1)} &
\colhead{(2)} &
\colhead{(3)} &
\colhead{(4)} &
\colhead{(5)} &
\colhead{(6)} &
\colhead{(7)} &
\colhead{(8)} &
\colhead{(9)} &
\colhead{(10)} &
\colhead{(11)} &
\colhead{(12)}
}
\startdata
S4 0636+68   &  6:42:04.2 & 67:58:36 & 16.6  &  3.180  & 90 & 11    &  47.28 & 131.9 &  36.6 &   133.2& 900 \\
S5 0836+71   &  8:41:24.3 & 70:53:42 & 16.5  &  2.172  & 70 & 16    &  46.81 &  66.7 & 341.2 & 10064.5& 600 \\
SBS 1116+603 & 11:19:14.3 & 60:04:57 & 17.5  &  2.646  & 85 & 15    &  46.92 & 147.2 &  62.7 &   632.7& 900 \\
SBS 1233+594 & 12:35:49.5 & 59:10:27 & 16.5  &  2.824  & 76 & 15    &  46.97 & 125.6 &  79.0 &     1.3& 600 \\
SBS 1425+606 & 14:26:56.2 & 60:25:51 & 16.5  &  3.192  & 90 & 21    &  47.43 &  78.6 &   2.4 &     4.6& 300 \\
HS 1700+6416 & 17:01:00.6 & 64:12:09 & 16.1  &  2.736  & 96 & 17    &  47.49 &  91.4 & 135.4 &     3.6& 300 \\
\hline
S5 0014+81   &  0:17:08.5 & 81:35:08 & 16.5  &  3.366  & 82 &\nodata&  47.56 &\nodata&\nodata&   493.3&\nodata\\
S5 0153+74   &  1:57:34.9 & 74:42:43 & 16.0  &  2.338  & 65 &\nodata&  46.85 &\nodata&\nodata& 12377.8&\nodata\\
TB 0933+733  &  9:37:48.8 & 73:01:58 & 17.3  &  2.528  & 86 &\nodata&  47.07 &\nodata&\nodata&     5.2&\nodata\\
HS 1946+7658 & 19:44:55.0 & 77:05:52 & 15.8  &  2.994  & 102&\nodata&  47.63 &\nodata&\nodata&     1.7&\nodata\\
S5 2017+74   & 20:17:13.1 & 74:40:48 & 18.1  &  2.187  & 77 &\nodata&  46.44 &\nodata&\nodata&  2765.8&\nodata
\enddata
\tablecomments{
Upper entries are objects monitored both photometrically and
spectrophotometrically.  Lower entries are objects monitored only
photometrically.
Col. (1) Object name. Cols. (2)--(3): Right ascension
and declination (2000) from NED. Col. (4): $V$ magnitude from
the Veron-Cetty \& Veron (1993) Catalogue. Col. (5): Redshift from
NED. Col. (6)--(7): Number of photometric and spectrophotometric
observing epochs. Col. (8): Log luminosity at 5100 \AA\ (luminosity
in units of \ergs) computed from the rest-frame flux density at
1450\,\AA\ of the averaged spectrum, using a power law of $f_\nu
\propto \nu^{-0.5}$, and correcting for Galactic absorption.
Cols. (9)--(10): Angular separation in arcseconds and position angle
in degrees of comparison star (see \S~\ref{specred}). Col. (11): Radio
(1.4 GHz)-to-optical (estimated at 4400\,\AA) flux ratio. Col. (12):
HET/LRS exposure time in seconds (two consecutive exposures with this
exposure time were obtained each visit).}
\end{deluxetable*}

The relations between AGN luminosity, BLR size, and black-hole mass
have been widely used for ``single-epoch'' estimates of black-hole
masses and accretion rates in studies addressing the issues of AGN
accretion history and black-hole growth (e.g., McLure \& Dunlop 2004;
Barger et al. 2005; Yu et al. 2005; Kollmeier et al. 2006). However,
these studies necessarily extrapolate the relations to luminosities
and redshifts well beyond the ranges in which they were measured
(e.g., McLure \& Jarvis 2002; Vestergaard 2002; Woo \& Urry 2002;
Vestergaard \& Peterson 2006), since the only measured relations are
in the luminosity range $10^{42}$--$10^{46}$\,\ergs\ and for redshift
$<0.3$. As a result, the conclusions of such studies depend on the
untested assumption that these extrapolations are valid. Although
{\it a posteriori} explanations of the physical plausibility of the
observed relations can be found, it is quite possible that subtle
or strong deviations from the relations occur at high luminosities
or redshifts (Netzer 2003). It is well known that some other
AGN properties, such as the optical-to-X-ray spectral slope and
emission-line equivalent widths, depend on luminosity (e.g., Baldwin
1977; Strateva et al. 2005; Steffen et al. 2006).  AGN outflows also
seem to be different between Seyfert galaxies (several hundred \kms
) and broad absorption line (BAL) quasars (several thousand to
tens of thousands \kms), and thus might depend on luminosity (Laor \&
Brandt 2002).  Actual reverberation measurements for AGN with high
luminosities and redshifts are therefore desirable.

However, reverberation mapping of high-luminosity quasars is an
ambitious undertaking. Quasars of the highest luminosities (with
bolometric luminosity, $L_{\rm bol} \approx 10^{47}$--$10^{48}$ \ergs
) are expected to harbor some of the most massive black holes known,
with $M_{\rm BH} \ga 10^9 M_{\sun}$. More massive black holes may
have slower continuum flux variations with smaller amplitudes. This
has been observed in X-rays (e.g., Lawrence \& Papadakis 1993;
Uttley et al. 2002; Markowitz et al. 2003; O'Neill et al. 2005),
and possibly in the UV-optical bands (e.g., Giveon et al. 1999; Cid
Fernandes et al. 2000; Vanden Berk et al. 2004).  The required
observing periods of high-luminosity quasars are also significantly
lengthened by cosmological time dilation, since such sources are
typically found at high redshifts ($z \ga 1$). On the other hand,
the ability to monitor high-$z$ objects in the rest-frame UV, in which
AGN variability amplitudes are routinely higher than in the optical,
can lead to better characterized continuum light curves.  The smaller
intrinsic variability amplitude of the continuum could result in
smaller flux-variability amplitudes for the emission lines, affecting
the ability to detect the time delay in the BLR response. Furthermore,
since the equivalent widths of high-ionization emission lines decrease
with increasing luminosity (the well-known `Baldwin Effect'; Baldwin
1977), fluxes of emission lines such as \ion{C}{4}$\lambda 1550$,
which are most suitable for such studies, are harder to measure. On
the other hand, the fluxes in the UV lines are generally higher than
those of the Balmer lines used in rest-frame optical studies. Finally,
high-redshift sources are fainter and hence more difficult to observe.
Probably due to all of these possible problems, no reverberation
measurements exist for AGNs with $L\gtrsim 10^{46}$\,\ergs , and
several attempts at such measurements have so far not been successful
(e.g., Welsh et al. 2000; Trevese et al. 2004; A. Marconi 2005,
private communication).

In view of the many unknowns and the opposing effects entering the
above discussion, and considering the importance of the subject, 
over a decade ago, we began a reverberation-mapping program aimed
at high-luminosity, high-redshift AGNs. In this paper, we describe
our program and present some initial results. In \S~2 we describe the
sample, the observations, and the data reduction. In \S~3 we perform
a time-series analysis for the AGN light curves and discuss our results,
with a summary in \S~4.
Throughout this paper we use the standard cosmology with
$H_{0}=70$\,km\,s$^{-1}$\,Mpc$^{-1}$, $\Omega_{M}=0.3$, and
$\Omega_{\Lambda}=0.7$ (Spergel et al. 2003; Riess et al. 2004).

\vglue0.5cm

\section{Sample, Observations, and Data Reduction}

\subsection{Sample Selection}

We selected our sample of 11 high-luminosity quasars in 1994 from
the Veron-Cetty \& Veron (1993) catalog.  The selection criteria were
observationally oriented: a high declination ($\delta \ga 60\degr$), so
as to make the quasars observable from the Northern hemisphere for most
of the year; observed magnitude $V\la18$, to permit the acquisition of
low-resolution, high signal-to-noise (S/N) spectra within a reasonable
observing time; and, since high-luminosity quasars are known only
at high redshifts, redshifts in the range $2<z<3.4$ that brings
the prominent broad UV emission lines (\ion{C}{4} and Ly$\alpha$)
into the optical range.  We chose the 11 objects with the brightest
apparent optical magnitudes that satisfy these criteria. The sample
is  listed in Table~\ref{sample}. The choice of bright quasars at high
redshift naturally favors quasars with high luminosity. The luminosity
range of our sample is $10^{46.4}\la\lambda L_\lambda(5100\,{\rm
\AA})\la10^{47.6}$ \ergs. This is an order of magnitude higher than
other AGNs with existing reverberation measurements.

\begin{figure}
\centerline{\includegraphics[width=8.5cm]{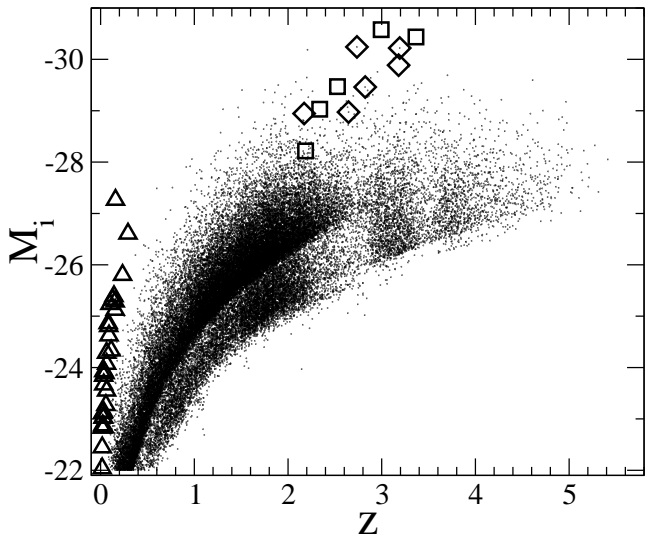}}
\vglue0.3cm
\caption{Absolute $i$ magnitude vs. redshift for all SDSS Data Release
3 quasars ($dots$). The six quasars monitored
at the HET are marked with diamonds, and the five objects from our
sample with photometric data only are marked with squares. AGNs with
reverberation-mapping data are marked with triangles. The objects in
our sample are clearly among the most luminous quasars and lie within
the primary epoch of black-hole growth for luminous quasars.
\label{Mizfig} }
\end{figure}

In Figure~\ref{Mizfig} we plot the absolute $i$ magnitude versus
redshift for all objects in the Sloan Digital Sky Survey (SDSS;
York et al. 2000) Data Release 3 (DR3; Abazajian et al. 2005) Quasar
Catalog (Schneider et al. 2005), compared to our sample and to other
objects with reverberation-mapping data. Since the SDSS DR3 quasar
catalog covers about 10\% of the sky, future surveys will not find
large numbers of quasars significantly more luminous than those in
it, and thus the objects in our sample are clearly among the most
luminous quasars in the Universe. Our sample includes HS\,1700+6416
and SBS\,1425+606 which are the two most-luminous quasars in
DR3. Our study is probing powerful quasars at redshifts near the
peak of their comoving number density (near $z\sim 2$), which is
the main era of black-hole growth for the most massive black holes
and galaxies. Figure~\ref{Mizfig} also shows that there are only two
objects with $M_i < -26$ that have a previously measured BLR size,
which illustrates the gap in luminosity between our sample and
previously studied AGNs.

\vglue0.5cm

The luminosities of the quasars in our sample are likely intrinsic,
rather than being affected by gravitational-lensing magnification.  To
our knowledge, none of the objects is a lensed quasar. In particular,
five of the quasars were imaged with the {\it Hubble Space Telescope}
(HST) by Maoz et al. (1993), who found no evidence for strong lensing.
Maoz et al. (1993) and other optical lensed-quasar surveys (e.g.,
Morgan et al. 2003) have consistently found a lensed fraction of $\sim
1$\% among luminous quasars, and thus it is not surprising that none
among our 11 objects is lensed.  We also note that none of our objects
is a BAL quasar, and none has significant UV
absorption lines.

Because the Veron-Cetty \& Veron (1993) catalog was a literature
compilation, our sample is not statistically complete, nor necessarily
representative of the luminous quasar population.  For example,
six of the quasars in our sample (i.e., about 55\%) are radio loud,
somewhat more than the $\sim 20$\% radio-loud fraction (i.e.,
2 objects) one would expect for a quasar population of this mean
redshift (2.8) and luminosity (Jiang et al. 2006). We
note that the incompleteness of AGN reverberation samples applies
to most of the studies that have derived BLR-size--mass--luminosity
relations. An exception is the sample of 17 PG quasars mapped by
Kaspi et al. (2000), which were an unbiased selection from the full,
flux-limited, color-selected Bright Quasar Survey (BQS; Schmidt \&
Green 1983; also see Maoz et al. 1994). Of course, the PG sample
itself suffers from some incompleteness; see Jester et al. (2005)
and references therein. Although there are no obvious biases that
are introduced into reverberation measurements by these samples
(e.g., radio-loud and radio-quiet objects seem to follow the same
BLR-size--luminosity relation; see the few radio-loud objects in the
Kaspi et al. 2000 sample and also Wu et al. 2004), it should be kept in
mind that such biases may, nonetheless, exist (see, e.g., Netzer 2003).

The observations and reductions for the current project follow the
same principles as those of similar projects carried out by our group
(e.g., Maoz et al. 1994; Netzer et al. 1996; Giveon et al. 1999; Kaspi
et al. 2000). We outline the main features in the following sections.

\subsection{Broad-Band Imaging}
\label{photom}

Imaging data have been obtained, starting in 1995, at the
Wise Observatory (WO; Kaspi et al. 1995) 1\,m telescope in
the Johnson-Cousins $B$ and $R$ bands, using a 1024$\times$1024
back-illuminated Tektronix CCD. The exposure times are 250~s in $R$
and 300~s in $B$. Since the targets have high declinations, they can
be observed from the WO for about 10 months a year, with observations
scheduled about once every month (based on the experience with the PG
quasars, Kaspi et al. 2000, and the expectation that the variability
timescales of high-luminosity quasars are longer, this sampling rate was
deemed sufficient). The data have been reduced using IRAF\footnote{IRAF
(Image Reduction and Analysis Facility) is distributed by the National
Optical Astronomy Observatories, which are operated by AURA, Inc,
under cooperative agreement with the National Science Foundation.}
procedures in the standard way. Broad-band light curves for the quasars
are produced by comparing their instrumental magnitudes to those of
constant-flux stars in the field (see, e.g., Netzer et al. 1996, for
details). The uncertainties on the photometric measurements include
the fluctuations due to photon statistics and the scatter (of 0.03
mag or less) in the measurement of the constant-flux stars used. The
typical total uncertainty is in the range of 0.01--0.05 mag depending
on object brightness and the observing conditions. The photometric
light curves provide a check on any continuum variability detected
in the spectroscopic observations and improve the sampling of the
continuum light curves. Furthermore, the photometric observations
began several years before the spectroscopic observations. In the
analysis, this measurement of the continuum behavior at early times
gives an enlarged baseline for the cross correlation between the line
and continuum light curves.

\subsection{Spectrophotometry}
\label{specred}

Spectrophotometric monitoring of six of the 11 quasars has been
carried out since 1999 at the 9\,m Hobby-Eberly Telescope (HET;
Ramsey et al. 1998; Ramsey et al. in preparation). The other
five quasars have declinations above the HET's high-declination limit
(71$\degr$41$\arcmin$). The operation of the HET in a queue-scheduled
mode is a key component of our observing program; it enables ideal
spreading of the observations throughout the observing season for
each object and ensures that observations will not be lost due to
poor weather. We have used the Low Resolution Spectrograph (LRS;
Hill et al. 1998) with a 600 line mm$^{-1}$ grating, a 10$\arcsec$
wide slit, and the GG385 order-separation filter. The spectral range
covered with this configuration is 4300--7300 \AA\, with a resolution
of about 20 \AA. Typical seeing in the HET data is in the range
1.5--3$\arcsec$. Each night the standard calibration images of bias,
sky-flat fields, and internal lamp-flats are obtained. Exposures of
Ne and Cd lamps are used for wavelength calibration (up to mid-2002
a Hg-Cd-Zn lamp was used). The objects in our sample are accessible
to the HET for about six months each year. During each such period,
we strive to obtain three observations separated by $\sim$2.5 months.
For each quasar, the spectrograph's focal plane is rotated to an
appropriate position angle so that a nearby comparison star (columns
[9] and [10] in Table~\ref{sample}) is included simultaneously in
the slit. This permits spectrophotometric calibration of the quasars
under non-photometric conditions. The wide slit reduces the effects
of atmospheric dispersion at the nonparallactic position angles,
as well as light losses due to guiding errors and poor seeing.

Observations typically consist of two consecutive exposures of the
quasar/star pair. Exposure times range from 300--900 seconds for each
exposure depending on object brightness, aiming for a S/N of $>20$
pixel$^{-1}$ in the continuum. The spectroscopic data are reduced
using standard IRAF routines. The images are bias and flat-field
corrected. The extraction width is typically 9.4$\arcsec$, and
wavelength calibration is applied to the data after the spectral
extraction. This procedure results in two spectra from each image:
one for the quasar and one for the comparison star. The consecutive
quasar/star flux ratios are then compared to test for systematic errors
in the observations and to identify and remove cosmic-ray events. The
ratio is usually reproducible to 0.5--4\% at all wavelengths;
observations with ratios larger than 5\% are discarded (this occurred
in about 7\% of the observations). We note a trend of improvement in
the HET observations in image quality and seeing over the years due to
improvement in its equipment and manner of operation. We have verified
that all our comparison stars are non-variable to within $\sim$2\%
by means of differential photometry with other stars in each field
(see \S\ref{photom}).

Spectra were calibrated to an absolute flux scale using observations
of spectrophotometric standard stars observed at several epochs.
The uncertainty in the calibration is $\sim$20\%, which is not shown
in the error bars in our light curves. The error bars reflect only the
measurement and differential uncertainties, which are of order 1--4\%.
These are comparable to other reverberation mapping studies which
have used the same observing technique (e.g., Kaspi et al. 1996a;
Collier et al, 1998; Kaspi et al. 2000).

\begin{figure}
\vglue0.5cm
\centerline{\includegraphics[width=8.5cm]{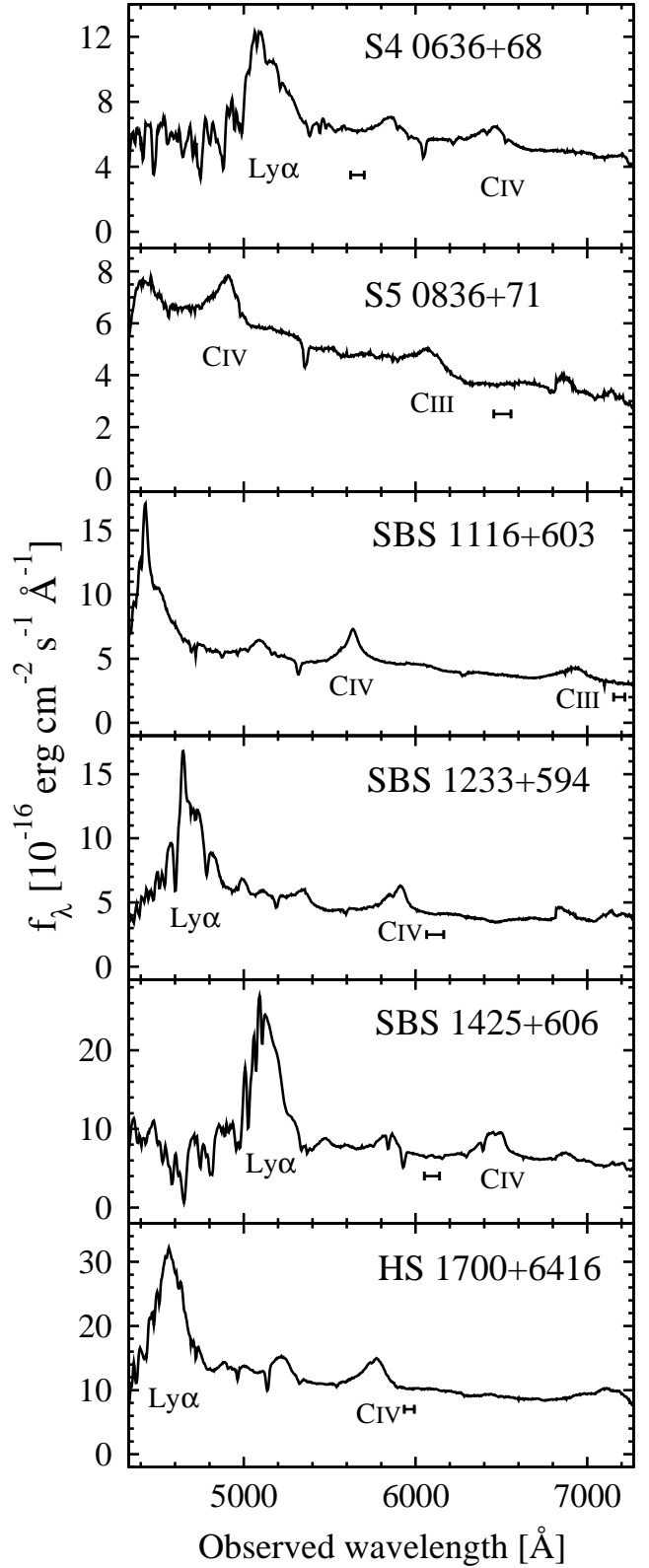}}
\vglue0.5cm
\caption{Mean spectra of the six quasars monitored with the HET. Each
spectrum contains roughly a dozen individual observations. The spectral
resolution is $\approx$ 20 \AA. The line and continuum bands discussed
in the text are marked in each panel.
\label{spectra} }
\end{figure}

For each quasar, we used all available spectra to produce an
average spectrum and a root-mean-square (rms) spectrum, defined as:
\begin{equation}
\sigma(\lambda)=\left\{{1\over (N-1)}\sum_{i=1}^N \left[f_i(\lambda)-\bar 
f(\lambda)\right]^2\right\}^{1/2},
\label{equ1}
\end{equation}
where the sum is taken over the $N$ spectra, and $\bar f(\lambda)$
is the average spectrum (Peterson et al. 1998a). The average spectra
of the six quasars in our sample are shown in Figure~\ref{spectra}.
We used the average and rms spectra to choose line-free spectral bands
suitable for setting the continuum underlying the emission lines,
and the wavelength limits for integrating the line fluxes. The line
and continuum fluxes were measured algorithmically for all epochs by
calculating the mean flux in the continuum bands, and summing the flux
above a straight line in $f_{\lambda}$ that connects the continuum
bands straddling each emission line. This process measures the total
emission in each line, i.e., the flux of the broad component of the
line together with its narrow component.  We note, however, that in the
high-ionization lines we monitor in luminous quasars, the narrow-line
contribution is negligible (e.g., Wills et al. 1993). In any event,
the narrow component is non-variable on the timescales probed here,
and would merely contribute a constant flux level to the light curves.

The uncertainty in the line flux was estimated
by propagating the uncertainty in the definition of the continuum
levels, determined from the standard deviation of the mean in the
continuum bands (of order 1--3\%). To this we added in quadrature our
estimate for the uncertainty in the calibration of the differential
spectrophotometry, which is of order 1--4\%.

\begin{deluxetable}{cc}
\tablecolumns{2}
\tablewidth{170pt}
\tablecaption{Integration Limits for Continuum Bands
\label{band}}
\tablehead{
\colhead{Object}        &
\colhead{Wavelength Range [\AA]}
} 
\startdata
S4 0636+68   &  5622--5701 \\
S5 0836+71   &  6456--6556 \\ 
SBS 1116+603 &  7154--7220 \\
SBS 1233+594 &  6065--6165 \\
SBS 1425+606 &  6051--6141 \\
HS 1700+6416 &  5935--5993 \\
\enddata
\end{deluxetable}

\vglue1.5cm

\subsection{Continuum and Emission-Line Light Curves}

As seen in all the quasar spectra in 
Figure~\ref{spectra}, the $R$ band ($\lambda\sim 6500$~\AA)
is dominated by continuum emission, with negligible contribution
from broad emission lines. In contrast, emission lines contribute of
order one-half the flux in the $B$ band for several of the quasars.
The variations in the $R$-band photometry are therefore determined
almost completely by continuum variations, and this broad-band light
curve can be merged with the spectroscopic continuum light curve,
to improve the sampling and duration of the continuum light curve.
The photometric and spectrophotometric light curves of each object
are intercalibrated by comparing all pairs of spectrophotometric and
photometric observations separated in time by less than 20 days. For
the data presented here, this produced $\sim10$ pairs of points per
object. A linear least-squares fit between the spectrophotometric
continuum fluxes and the photometric fluxes (the instrumental
$R$-magnitudes were translated to fluxes) was then used to merge
the two light curves. A clear relation was always apparent in this
fitting and a linear model was an adequate representation, indicating
the continuum light curves from the spectroscopic and the photometric
 observations are in good agreement.

\begin{figure*}
\centerline{\includegraphics[width=18cm]{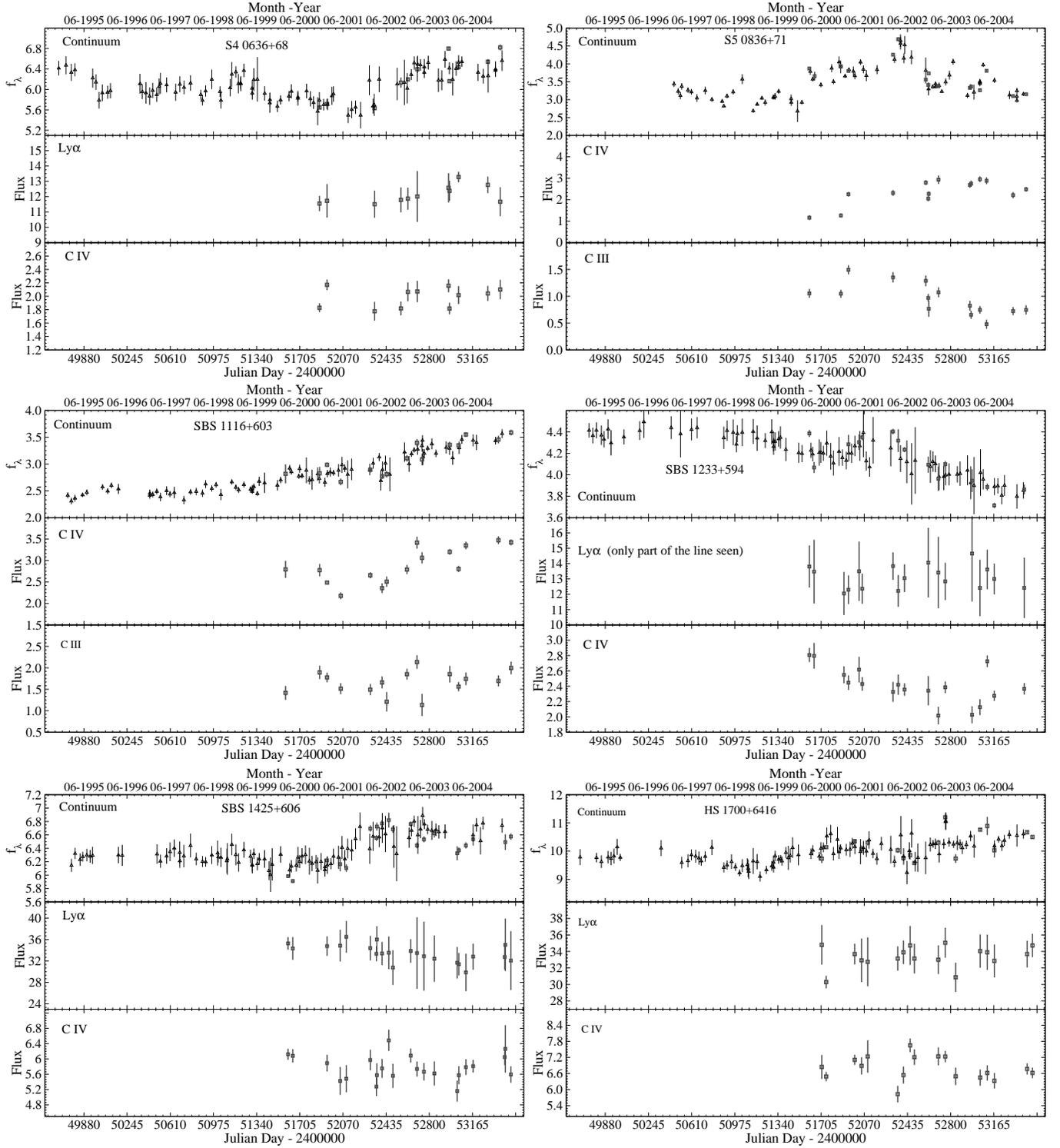}}
\vglue0.5cm
\caption{Light curves for the six quasars which have been monitored at
the HET. Squares are spectrophotometric data from the HET. Triangles
are photometric data from WO. Time is given in Julian Day ({\it
bottom}) and UT date ({\it top}).  Continuum flux densities,
$f_\lambda$, are given in units of $10^{-16}$\,\ergcmsA\ and
emission-line fluxes are given in units of $10^{-14}$\,\ergcms.
\label{lc} }
\end{figure*}

\begin{figure}
\vglue0.2cm
\centerline{\includegraphics[width=8.5cm]{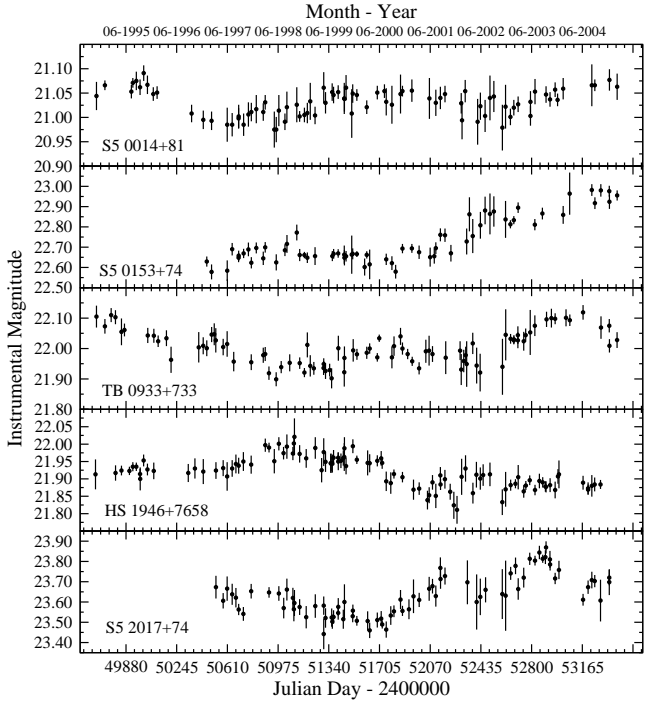}}
\caption{$R$-band light curves for the five quasars which have been
monitored photometrically at the WO. Magnitudes are
given in arbitrary units. Time is given in Julian Day ({\it bottom})
and UT date ({\it top}).
\label{phot_lc} }
\end{figure}

For each object, the total number of photometric and spectrophotometric
observations is given in Table~\ref{sample}, columns (6) \& (7),
respectively (the sum of the two columns gives the total number of
points in the continuum light curve). Note that for particular emission
lines in several objects, some data points are missing because of
insufficient wavelength coverage or low S/N. Light curves for the
six quasars monitored at the HET are presented in Figure~\ref{lc}.
The different bands used for the flux measurements in Figure~\ref{lc}
are listed in Table~\ref{band} for each object (wavelengths are
given in the observed frame). $R$-band light curves for the five
objects monitored photometrically at the WO are presented in
Figure~\ref{phot_lc}.

\section{Time Series Analysis}

\subsection{Continuum and Line Variability}

In Table~\ref{vartab}, we list variability measures for all light
curves presented above. Column (1) lists the object name and column
(2) the particular light curve. In columns (3), (4), and (5) we
list the mean ($\bar{f}$), rms ($\sigma$), and the mean uncertainty
($\delta$) of all data points in the light curves, respectively,
in the appropriate units for each light curve. In column (6)
we list the $\chi^{2}_{\nu}$ obtained by fitting a constant to
the light curve, and in column (7) we list P$(\chi^{2}|{\nu})$,
the probability to get such $\chi^{2}$ if there were intrinsically
no variability. This test is used to determine whether the light
curve is consistent with a constant flux to a significance level of
95\%. If the light curve passed the test, we list for it in column
(8) the intrinsic normalized variability measure, $\sigma_{N} =
100\sqrt{\sigma^{2}-\delta^{2}}/\bar{f}$ (used by Kaspi et al. 2000),
and in column (9) we list the fractional variability amplitude,
$F_{var}$, and its uncertainty (Rodriguez-Pascual et al. 1997;
Edelson et al. 2002).

In general, all continuum light curves show variations of 10--70\%
measured relative to the minimum flux.  The mean intrinsic normalized
variability measure (column 8 of Table~\ref{vartab}) is 5.8\% and
the median is 4.1\%. The points that appear to deviate strongly from
the mean flux are generally due to unreliable measurements and have
relatively large uncertainties. A possible exception is one particular
deviation in the continuum light curve of HS\,1700+6416 around
JD$\sim$2452757 (2003 April 27) where the object seems to brighten
(both in the spectroscopic and in the photometric measurements)
by about 10\% for $\sim 15$ days.  This object was monitored
also by Reimers et al. (2005), who found that on 2003 May 14--18
HS\,1700+6416 was 20\% brighter in the $R$ band than in March 1995
and July 1998. This agrees with our measurements at the same epochs
and suggests that the above deviation in the light curve is real.

It is instructive to compare the variability characteristics of
our sample to those of the low-redshift PG quasar sample (Kaspi
et al. 2000).  However, one needs to consider the differences in
the rest-frame monitoring periods of the two samples. While the
continuum light curves of the high-luminosity quasars span over
10 yr in the observed frame, this period corresponds to only about
2--3 yr in their rest frames. We have therefore analyzed the first
3 yr period in the PG quasar light curves (which span over 7.5 yr).
The resulting variability measured for the PG quasar sample, 25--125\%,
is about double the 10--70\% variability range of the high-luminosity
quasars. Also, the mean and rms of $\sigma_{N}$ for the PG quasars
(in the first 3 years) is $<$$\sigma_{N}$$>$=(10.4$\pm$4.5)\%
about twice that of the high-luminosity quasars which have
$<$$\sigma_{N}$$>$=(5.8$\pm$3.9)\%.  The lower rest-frame variability
we measure in the continuum for the current sample is probably a
manifestation of the general trend that high-luminosity AGNs have
longer variability timescales (e.g., Vanden Berk et al. 2004),
perhaps as a result of their higher black hole masses.

Proceeding to the emission-line light curves, the variability
measures listed in Table~\ref{vartab} show that none of the four
Ly$\alpha$ light curves shows significant variability. In contrast,
the two \ion{C}{3}]$\lambda$1909\footnote{The \ion{C}{3}] complex we measure includes up to $\sim$50\% flux contribution from \ion{Si}{3}]$\lambda$1892 and
\ion{Fe}{3} multiplets.} light curves and all six \ion{C}{4} light curves
show significant variability, with $F_{var}$ significant at the
$\ga$3$\sigma$ level. We note that the variability measures of these
emission-line light curves are comparable to, or even greater than,
those of their corresponding continuum light curves.  There are
few previous AGN UV data sets with which to compare these possible
trends. The only quasar with UV variability data of similar quality is
3C273. Interestingly, Ulrich et al. (1993) noted the non-variability
of Ly$\alpha$ in this object, at a level of $<5$\%, over a period
of 15 years, despite factor-of-two variations in the continuum
during the same period (there are no data for the \ion{C}{4} and
\ion{C}{3}] lines during that time).  In contrast, Seyfert galaxies
seem to show comparable variation amplitudes in the UV continuum
and in the Ly$\alpha$, \ion{C}{4}, and \ion{C}{3}] lines, though
the line-variation amplitudes tend to be somewhat smaller than the
continuum variations. This has been seen in NGC~5548 (Clavel et
al. 1991; Korista et al. 1995), NGC~7469 (Wanders et al. 1997),
NGC~3783 (Reichert et al. 1994), Fairall 9 (Rodriguez-Pascual et
al. 1997) and 3C390.3 (O'Brien et al. 1998).

The indication  that, in quasars, the \ion{C}{4}, \ion{C}{3}], and
Ly$\alpha$ lines show different variability trends may be related
to the fact that each line is driven by a different ionizing
continuum. This could also explain the different variability
amplitudes, e.g., if the Ly$\alpha$ line is driven by a weakly
varying UV continuum, while the \ion{C}{4} is controlled by a
more strongly varying extreme-UV to X-ray continuum.  Indeed, the
continuum variability amplitudes of AGN are observed to grow with
decreasing wavelength, from optical through the UV and to the X-rays,
and the variations are sometimes poorly correlated with each other in
different bands (e.g., Maoz et al. 2002; Shemmer et al. 2003). The
differences in response behaviors among quasars and Seyferts would
then imply that the degree to which continuum variations grow with
decreasing wavelength depends on luminosity.  However, the current
data are still sparse, and the above trends require confirmation.

Comparison of the variability in the emission-line light curves between
the PG sample and the present high-luminosity sample is complicated by
the fact that, in the PG sample, the Balmer lines were monitored, while
for the high-luminosity quasars we monitor UV lines. Furthermore, the
high-luminosity quasar lines were monitored for only half the period
over which the continuum was monitored, i.e., approximately 5 yr,
which is about 1.2--1.5 yr in the quasar rest frames.  Analyzing only
the first 1.5 yr of the Balmer-line light curves of the PG quasars,
we find that about one-third of the lines did not vary; most of
the non-varying light curves are for the H$\gamma$ line which has
lower EW than the H$\alpha$ and H$\beta$ lines. If, for the PG-quasar
emission lines, we average the variability measures of only H$\alpha$
and H$\beta$, we find $<$$\sigma_{N}$$>$=(7.5$\pm$6.2)\% (mean and
rms), compared to the average variability measure of the \ion{C}{3}]
and \ion{C}{4} lines of the high-luminosity quasars in the current
sample, $<$$\sigma_{N}$$>$=(12.8$\pm$9.4)\%.  It therefore appears
that the UV lines may be more responsive than the Balmer lines to
continuum variations, which could compensate for the lower continuum
variability amplitudes at high luminosities, and contribute to the
success of our experiment.  Indeed, the high ratio of emission-line
to continuum-variation amplitudes in the high-luminosity sample
is in contrast to the common behavior in the PG quasars which show
less variability in the Balmer lines than in the continuum. [Some
low-luminosity AGNs, e.g., NGC\,4051 (Shemmer et al. 2003), do show
larger variability amplitudes in the Balmer lines than in the optical
continuum.] However, due to the limited rest-frame period considered
here, we consider this result tentative.

We do not see any strong FWHM or line-profile variations in our
sample of high-luminosity quasars. However, at the low resolution of
our spectra (about 1200\,\kms\ at 5000\,\AA ), our ability to detect
such variations is limited to dramatic changes only.

\begin{deluxetable*}{llrccrccc}
\tablecolumns{9}
\tabletypesize{\scriptsize}
\tablewidth{400pt}
\tablecaption{Variability Measures
\label{vartab}}
\tablehead{
\colhead{Object}        &
\colhead{Light Curve}   &
\colhead{Mean\tablenotemark{a}}     &
\colhead{RMS\tablenotemark{a}} &
\colhead{Mean Uncertainty\tablenotemark{a}} &
\colhead{$\chi^{2}_{\nu}$} &
\colhead{P$(\chi^{2}|{\nu})$\tablenotemark{b}} &
\colhead{$\sigma_N$} &
\colhead{$F_{var}$} \\
\colhead{(1)} &
\colhead{(2)} &
\colhead{(3)} &
\colhead{(4)} &
\colhead{(5)} &
\colhead{(6)} &
\colhead{(7)} &
\colhead{(8)} &
\colhead{(9)} 
} 
\startdata
S4 0636+68 & continuum &  6.09 &  0.29 &  0.16 & 19.76 & 0 &  4.10 &$0.039\pm0.004$ \\ 
           & Ly$\alpha$& 12.10 &  0.57 &  0.83 &  1.31 & 0.22 &\nodata&     \nodata \\ 
           & \ion{C}{4}&  1.99 &  0.15 &  0.11 & 2.40  & $7.6\times10^{-3}$ &  4.70 &$0.045\pm0.027$ \\ 
S5 0836+71 & continuum &  3.48 &  0.45 &  0.09 & 80.03 & 0 & 12.53 &$0.124\pm0.010$ \\ 
           & \ion{C}{4}&  2.36 &  0.57 &  0.13 & 22.49 & 0 & 23.37 &$0.233\pm0.047$ \\ 
           & \ion{C}{3}]& 0.94 &  0.29 &  0.09 & 11.14 & 0 & 29.36 &$0.293\pm0.061$ \\ 
SBS 1116+603& continuum&  2.85 &  0.35 &  0.09 & 30.49 & 0 & 11.80 &$0.117\pm0.009$ \\ 
           & \ion{C}{4}&  2.89 &  0.42 &  0.10 & 26.17 & 0 & 14.02 &$0.140\pm0.027$ \\ 
           & \ion{C}{3}]& 1.66 &  0.28 &  0.16 &  2.46 & $1.7\times10^{-3}$ & 13.92 &$0.137\pm0.037$ \\ 
SBS 1233+594& continuum&  4.19 &  0.19 &  0.11 & 10.46 & 0 &  3.65 &$0.034\pm0.004$ \\ 
           & Ly$\alpha$& 13.12 &  0.76 &  1.57 &  0.27 & 1.00 &\nodata&     \nodata \\ 
           & \ion{C}{4}&  2.41 &  0.24 &  0.11 &  5.06 & 0 &  8.62 &$0.085\pm0.019$ \\ 
SBS 1425+606& continuum&  6.38 &  0.22 &  0.12 & 11.25 & 0 &  2.86 &$0.027\pm0.003$ \\ 
           & Ly$\alpha$& 33.42 &  1.67 &  3.14 &  0.47 & 0.99 &\nodata&     \nodata \\ 
           & \ion{C}{4}&  5.77 &  0.32 &  0.26 &  1.62 & 0.036 &  3.31 &$0.028\pm0.017$ \\ 
HS 1700+6416& continuum& 10.00 &  0.41 &  0.23 & 10.87 & 0 &  3.37 &$0.032\pm0.003$ \\ 
           & Ly$\alpha$& 33.38 &  1.29 &  1.86 &  1.17 & 0.28 &\nodata&     \nodata \\  
           & \ion{C}{4}&  6.80 &  0.45 &  0.30 &  2.64 & $3.6\times10^{-4}$ &  4.87 &$0.046\pm0.016$ \\ 
\hline
S5 0014+81 &     $R$   & 21.030&  0.027&  0.024&  1.93 & $1.1\times10^{-6}$ &  1.21 &$0.008\pm0.006$ \\ 
S5 0153+74 &     $R$   & 22.731&  0.113&  0.039& 14.13 & 0 &  9.34 &$0.092\pm0.010$ \\ 
TB 0933+733&     $R$   & 22.003&  0.055&  0.031&  5.11 & 0 &  4.19 &$0.039\pm0.005$ \\ 
HS 1946+7658&    $R$   & 21.919&  0.043&  0.025&  3.66 & 0 &  3.21 &$0.031\pm0.004$ \\ 
S5 2017+74 &     $R$   & 23.631&  0.100&  0.049&  9.36 & 0 &  7.86 &$0.075\pm0.009$ \\ 
\enddata
\tablecomments{No data in columns (8) \& (9) mean that no significant
variability was detected in this light curve. In particular none of
the Ly$\alpha$ light curves shows significant variability}
\tablenotetext{a}{The units are
10$^{-16}$~ergs\,cm$^{-2}$\,s$^{-1}$\,\AA$^{-1}$ for the continuum
light curves, 10$^{-14}$~ergs\,cm$^{-2}$\,s$^{-1}$ for the emission-ine light
curves, and instrumental magnitude for the $R$-band light curves.}
\tablenotetext{b}{Probabilities smaller than $10^{-9}$
are listed as zero.}
\end{deluxetable*}

\subsection{Line-Continuum Cross Correlations}

% \vglue-0.4cm
The ultimate objective of our program is to detect and measure a
time delay between the continuum and the line-flux variations in
high-luminosity AGNs. The significant continuum and line variations
that we have observed during a decade demonstrate that, at least in
principle, such a measurement may be feasible. However, an actual
estimate of a delay requires: (1) high variability amplitude in both
the line and the continuum light curves; and (2) light curves with at
least a few, reasonably sampled, large-amplitude ``events'', i.e.,
changes of sign in slope, that permit unambiguous matching between
continuum and line variations. Examining the light curves of the six
quasars with emission-line data at the current stage of our project,
all but one currently suffer from either low variability amplitude
in the emission-line light curves or monotonically increasing or
decreasing continuum light curves. In four objects the \ion{C}{4} light
curve tracks the continuum's monotonic variation (SBS\,1116+603,
SBS\,1233+594, SBS1425+606, HS\,1700+6416). We note that the
monotonic variation trend for these objects is unlikely to continue
indefinitely. We expect that at some point, there will be a ``break''
in each of these cases, permitting the eventual measurement of a lag.
However, the one current exception to this state of affairs is
S5\,0836+71; although the data for this quasar are still not ideal in
terms of the criteria above, they do allow a preliminary measurement
of the emission-line to continuum lag. The continuum light curve for
this object displays a general rise until June 2002, followed by a
sharp drop in flux.  This light curve has largest variation among
all our monitored quasars, and the light curves of both \ion{C}{4}
and \ion{C}{3}] display the largest variations among all the emission
line light curves. Both emission-line light curves seem to follow the
general trend of the continuum light curves, although the \ion{C}{4}
light curve seems to have a much larger time lag than the \ion{C}{3}]
light curve.

\begin{figure}
\centerline{\includegraphics[width=8.5cm]{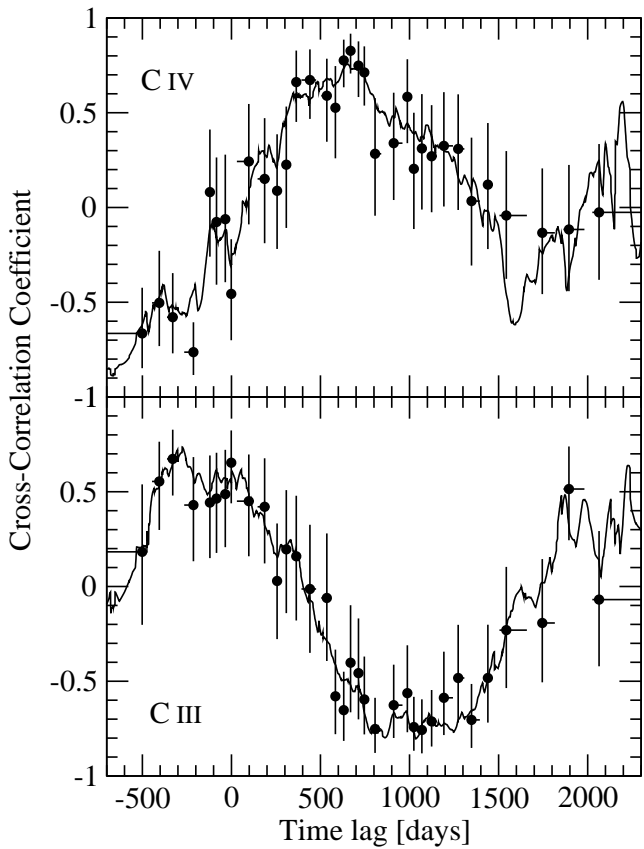}}
\caption{Cross-correlation functions, ICCF ({\it solid curve}) and ZDCF
({\it circles with error bars}), between the continuum and the emission-line
(\ion{C}{4} - {\it top}, \ion{C}{3}] - {\it bottom}) light curves of
S5\,0836+71 from Fig.~\ref{lc}.
\label{ccf} }
\end{figure}

To quantify the time lag, we use two methods for cross-correlating the
line and continuum light curves. The first method is the interpolated
cross-correlation function (ICCF), as implemented by White \&
Peterson (1994; see also the review by Gaskell 1994). The second
method is the $z$-transformed discrete correlation function (ZDCF)
of Alexander (1997).  The two methods yield similar results for the
current data, and we will use only the ICCF results in the following
analyses. The results of the cross-correlation analysis are presented
in Figure~\ref{ccf}.

The CCFs for S5\,0836+71 show enough structure to permit estimation
of a time lag. The clearest case is the correlation between the
\ion{C}{4} line and the continuum. To quantify the possible time
lag and its uncertainty, we use the the model-independent FR/RSS
Monte Carlo method of Peterson et al. (1998b; 2005). In this method,
each Monte Carlo simulation is composed of two parts. The first is a
``random subset selection'' (RSS) procedure which consists of randomly
drawing, with replacement, from a light curve of $N$ points a new
sample of $N$ points, while preserving the temporal order. The second
part is ``flux randomization'' (FR) in which the observed fluxes
are altered by random Gaussian deviates scaled to the uncertainty
ascribed to each point. The two resampled and altered time series are
cross-correlated using the ICCF method, and the centroid of the CCF
is computed. We used $\sim$10000 Monte Carlo realizations to build
up a cross-correlation centroid distribution (CCCD; e.g., Maoz \&
Netzer 1989). The mean of the distribution is taken to be the time
lag and the uncertainty is determined as the range which contains 68\%
of the Monte Carlo realizations in the CCCD, and thus would correspond
to 1$\sigma$ uncertainties for a normal distribution.

We find the time lag between the \ion{C}{4} line and the continuum
of S5\,0836+71 to be $595^{+86}_{-110}$ days, or $188^{+27}_{-37}$
days in the quasar rest frame. However, 17\% of the CCCD simulations
failed to produce significant centroid measurements. This may indicate
that the quality of the data render this analysis premature. The cross
correlation of the \ion{C}{3}] light curve with the continuum yields a
peak around zero time lag. A formal CCCD gives $-152^{+199}_{-182}$
days, which is $-48^{+63}_{-57}$ days in the rest frame, but the CCCD
function is not simple, with several peaks within this uncertainty
range. We note that the only AGNs in which (rather uncertain) 
\ion{C}{3}] time lags
have been estimated to date are NGC\,5548 (Peterson et al. 2004, and
references therein) and NGC\,4151 (Metzroth et al. 2006).

\begin{figure}
\centerline{\includegraphics[width=8.5cm]{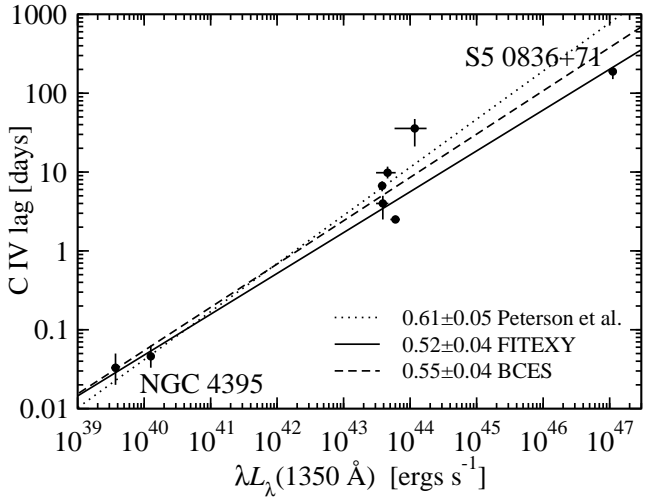}}
\caption{Size-luminosity relationship based on the \ion{C}{4}
$\lambda$1549 emission line and the UV continuum. Data
(circles) are reproduced from Peterson et al. (2005), to which we
add S5\,0836+71 from the current study. The new data point deviates
somewhat from the best-fit relation found by Peterson et al. (dotted
line). Our own linear fits to the data are marked with a solid line
for the FITEXY method and a dashed line for the BCES method.
The power-law slopes of the three relations are indicated. 
\label{rvsluv} }
\end{figure}

Until recently, only four AGNs had measured \ion{C}{4} reverberation
time lags: NGC\,3783, NGC\,5548, NGC\,7469, and 3C\,390.3 (see
Peterson et al. 2004, for a summary). Peterson et al. (2005) have
now measured the \ion{C}{4} time lag also for the low-luminosity
Seyfert galaxy NGC\,4395, which is four orders of magnitude lower in
luminosity than those four AGNs, permitting a first estimate of the BLR
radius-luminosity relationship for the \ion{C}{4}-emitting region. Our
preliminary determination of the \ion{C}{4} time lag for S5\,0836+71
allows us to extend this relation to seven orders of magnitude in
luminosity. In Figure~\ref{rvsluv}, we show the data as presented
by Peterson et al. (2005), as well as their adopted best-fit relation
(dotted line), to which we have added the data point for S5\,0836+71.
The rest-frame UV luminosity of S5\,0836+71 has been computed from the
mean flux during our observations. We have corrected the luminosity
for a Galactic extinction of $A_V=0.101$ from Schlegel et al. (1998)
using the extinction curve of Cardelli et al. (1989). We find the
luminosity to be $\lambda L_\lambda$(1350\AA ) = $(1.12\pm0.16)\times
10^{47}$ \ergs . Our new data point deviates from the extrapolation of
the fit by Peterson et al. (2005), which would predict a rest-frame
time delay for S5\,0836+71 of 372--1865 days.  We note that, due to the
current length of our program, we are sensitive to time lags of up
to about 1000 days in the observer frame.

Most previous reverberation-mapping experiments have measured the
Balmer emission lines, in particular H$\beta$. For the four AGNs
above, H$\beta$ reverberation measurements exist, in addition to
the \ion{C}{4} measurements (Peterson et al. 2004). The scaling
between the BLR sizes of the two lines is important since it is the
means by which the BLR size from \ion{C}{4} reverberation mapping of
high-luminosity quasars can be compared to the BLR size from H$\beta$
reverberation mapping of Seyferts and low-luminosity quasars. For
NGC\,3783, NGC\,5548, and NGC\,7469 the H$\beta$ delays are factors of
2.55, 1.98, and 1.8, respectively (which average to $\sim 2$) larger
than the \ion{C}{4} delays, but for 3C\,390.3 the \ion{C}{4} delay is
larger than the H$\beta$ delay by a factor of 1.5. Thus, the relation
between the \ion{C}{4} and H$\beta$ BLR sizes is still unclear, due to
the small number of objects with reverberation measurements in both
lines, and the small spread in luminosity among these four AGNs.
BLR stratification suggests that the \ion{C}{4} emission should
arise interior to the H$\beta$ emission. This is confirmed by the
three Seyfert 1s but not by the measurements of 3C\,390.3. Also, the
very small spread of these four AGNs in luminosity does not allow
a meaningful scaling of the H$\beta$ BLR size to the \ion{C}{4}
BLR size; broadening the luminosity range is essential.

We  have performed a linear regression for the \ion{C}{4}
BLR-size--luminosity relation using all current data points in
Figure~\ref{rvsluv}. We have used the two methods which are
described in detail in Kaspi et al. (2005): (1) the FITEXY from
Press et al. (1992, p. 660) which is based on an iterative process to
minimize $\chi^2$ and in which we adopted the Tremaine et al. (2002)
procedure to account for the intrinsic scatter in the data; (2) the
bivariate correlated errors and intrinsic scatter (BCES) regression
method of Akritas \& Bershady (1996). Both methods take into account
the uncertainty in both coordinates and the intrinsic scatter around
a straight line. The fit for the data we find from the FITEXY method
(solid line in Figure~\ref{rvsluv}) is
\begin{equation}
\frac{R_{BLR}}{10 {\rm \ lt\,days}}= (0.17\pm0.04)\left(\frac{\lambda
L_{\lambda}(1350\,{\mbox{\AA}} )}{\rm 10^{43} \ ergs\,s^{-1}}
\right)^{0.52\pm0.04} ,
\end{equation}
and for the BCES method (dashed line in Figure~\ref{rvsluv}) it is
\begin{equation}
\frac{R_{BLR}}{10 {\rm \ lt\,days}}= (0.24\pm0.06)\left(\frac{\lambda L_{\lambda}(1350\,{\mbox{\AA}} )}{\rm 10^{43} \ ergs\,s^{-1}} \right)^{0.55\pm0.04} .
\end{equation}
The Pearson correlation coefficient is 0.97 with significance level
of $7\times 10^{-5}$ and the Spearman rank-order correlation is 0.83
with a significance level of 0.01. The intrinsic scatter we find using
the Tremaine et al. (2002) procedure is 54\%. The slope we find for
the relation between the \ion{C}{4} BLR-size and UV luminosity is
very close to the one found by Kaspi et al. (2005) between H$\beta$
BLR-size and the UV luminosity ($\sim0.55$).

The mean FWHM of the \ion{C}{4} line measured from the mean spectrum of
S5\,0836+71 is about 9700\,\kms . Using Eq.~5 of Kaspi et al. (2000)
and the time lag of 188 days, we estimate the central mass of
S5\,0836+71 at about $2.6\times10^{9}$M$_{\sun}$. This is the highest
mass directly measured for the black hole in an AGN using reverberation
mapping. 3C\,273 (=PG\,1226+023), the quasar with the highest directly
measured mass so far, has a mass of $8.9\times10^{8}$M$_{\sun}$,
$\lambda L_\lambda$(1350\,\AA )=2.0$\times10^{46}$ \ergs , and
$\lambda L_\lambda$(5100\,\AA )=9.1$\times10^{45}$ \ergs . Thus,
S5\,0836+71 has a factor three higher mass and a factor $\sim 6$
higher luminosity than 3C\,273.

The central mass of S5\,0836+71 corresponds to an Eddington luminosity
of $L_{\rm Edd}=3.9\times10^{47}$\,\ergs\ (for solar-abundance gas
$L_{\rm Edd}\approx 1.5\times10^{38} M/M_{\sun}$, e.g., Shaviv 1998).
Using $\lambda L_\lambda$(5100\,\AA )=6.4$\times10^{46}$ \ergs\
for S5\,0836+71 and a bolometric correction of 5.6 from Marconi
et al. (2004, Eq. 21) we find a bolometric luminosity, $L_{\rm
bol}=3.6\times10^{47}$ \ergs . Thus, this object is accreting close
to its Eddington luminosity, with $L_{\rm bol}/L_{\rm Edd} \approx
0.9$. We note that using Eq. 7 from Vestergaard \& Peterson (2006) to
estimate the black-hole mass from the UV luminosity and the \ion{C}{4}
FWHM, one obtains $M_{\rm BH}=1.8\times10^{10}$M$_{\sun}$, which is 7
times higher than our tentative measurement from reverberation mapping.

The high Eddington ratio that we obtain is similar to that found for
many other quasars (e.g., Shemmer et al. 2004; Baskin \& Laor 2005).
However, we note that there are several caveats which could influence
our result. The first is the tentative nature of our measurement of
the time lag for the \ion{C}{4} line. One should keep in mind the
possibility that the correlation and delay that we have measured in
S5\,0836+71 might be a chance coincidence, since it is the single
object in which we have found a correlation, out of six objects where
we have searched for one.  Another concern is our measured luminosity
of the object. Since S5\,0836+71 is the radio-loudest object in
our spectrophotometric sample it could be that a jet-linked, beamed
optical continuum is influencing the variability and the continuum-flux
measurement. While a beamed component might influence the variability
on small timescales (the continuum light curve does show variability
on timescales of months) the \ion{C}{4} line light curve is smooth and
its response is to the long-timescale variations seen in the continuum
light curve. We have also checked the \ion{C}{4} equivalent widths
(EWs) of the six quasars monitored at the HET, of which three are
radio loud and three are radio quiet. For the radio-loud quasars, the
observed EWs are 33, 68, and 101\,\AA\ while for the radio-quiet ones
the EWs are 57, 68, and 90\,\AA. Thus, both classes have comparable
\ion{C}{4} EWs, and there is no clear difference that might indicate
that there is an additional contribution to the UV continuum luminosity
of the radio-loud quasars.

\section{Summary}

We have presented first results from our long-term program to
monitor high-luminosity, high-redshift quasars, with the objective
of eventually measuring the sizes of their broad-line regions and
the masses of their central black holes.  All 11 quasars in our
photometrically monitored sample show continuum variability of 10--70\%
over a rest-frame period of 2--3 years. Compared to previous studies of
lower luminosity AGNs, the continuum variability amplitude of higher
luminosity quasars over the same rest-frame interval appears to be
smaller in the current high-luminosity sample.  Six of the 11 quasars
have been monitored spectroscopically over the past $\sim$5 years
(1.2--1.5 yr in the quasar rest frames) using the HET.  None of the
quasars show Ly$\alpha$ variations, at a level of $<7\%$, and this
non-variability may be a generic feature of high-luminosity AGN.
However, we do detect variable \ion{C}{3}] and \ion{C}{4} broad line
fluxes, whenever these lines are in our observed spectral range. In
several cases, these lines track the continuum variations in the
same quasar.  The variations in the broad \ion{C}{3}] and \ion{C}{4}
lines are higher than Balmer-line variations in low-luminosity quasars
over the same rest-frame interval, but this result is tentative,
in view of the still-limited spectroscopic monitoring period.

In one object, S5\,0836+71, we tentatively measure the
\ion{C}{4} rest-frame time lag behind the continuum emission to
be $188^{+27}_{-37}$ days. This is lower than expected from an
extrapolation of the \ion{C}{4} BLR-size--luminosity relation measured
at lower luminosities. Our measurement permits tracing this relation
over seven orders of magnitude in AGN luminosity. From this time
lag we estimate a black hole mass of $2.6\times10^{9}$ M$_{\sun}$
in S5\,0836+71, the largest ever measured by this technique.

Our results to date demonstrate that reverberation mapping of
high-redshift, high-luminosity, quasars may be feasible.  Based on
our past experience at quasar reverberation (Maoz et al. 1994;
Kaspi et al. 1996b, 2000), we hope that with $\approx 5$ more years
of similar data we will be able to establish a reliable BLR size for
S5\,0836+71, and perhaps for several other objects in our sample. The
unpredictability of quasar light curves, together with our lack of
knowledge about the response amplitudes and timescales of the UV
lines, makes it difficult to estimate our chances of success. If we
do succeed, the combined reverberation results from this program and
those from a program being carried out on very low-luminosity AGNs
will cover the {\it entire} AGN luminosity range and much of cosmic
time, including the peak era of black-hole growth for the most massive
black holes and galaxies. This will enable a direct measurement of
the black-hole mass in these AGNs and a more reliable indirect mass
measurement for {\it all} AGNs.

\acknowledgments

%\vglue0.4cm
We are grateful to the staffs of WO and HET for their great help in
carrying out this long-term program. Special thanks go to John Dann,
Ezra Mashal, and Sami Ben-Gigi of the WO and to Gary Hill of the
HET for devoted technical support of this project through the years.
We thank Ari Laor for helpful discussions, and Mike Eracleous and Larry
Ramsey for their support.  We thank the anonymous referee for several
valuable suggestions.  The Hobby-Eberly Telescope (HET) is a joint
project of the University of Texas at Austin, the Pennsylvania State
University, Stanford University, Ludwig-Maximillians-Universit\"at
M\"unchen, and Georg-August-Universit\"at G\"ottingen. The HET
is named in honor of its principal benefactors, William P. Hobby
and Robert E. Eberly. The Marcario Low-Resolution Spectrograph is
named for Mike Marcario of High Lonesome Optics, who fabricated
several optics for the instrument but died before its completion;
it is a joint project of the Hobby-Eberly Telescope partnership
and the Instituto de Astronom\'{\i}a de la Universidad Nacional
Aut\'onoma de M\'exico.  We gratefully acknowledge the financial
support of the Colton Foundation at Tel-Aviv University (S.~K.), the
Zeff Fellowship at the Technion (S.~K.), NASA LTSA grant NAG5-13035
(W.~N.~B., D.~P.~S., O.~S.), the Israel Science Foundation grant 232/03
(H.~N.), and NSF grant AST03-07582 (D.~P.~S.).  This research has made
use of the NASA/IPAC Extragalactic Database (NED) which is operated
by the Jet Propulsion Laboratory, California Institute of Technology,
under contract with the National Aeronautics and Space Administration.
Funding for the creation and distribution of the SDSS and SDSS-II has
been provided by the Alfred P. Sloan Foundation, the Participating
Institutions, the National Science Foundation, the U.S. Department
of Energy, the National Aeronautics and Space Administration, the
Japanese Monbukagakusho, the Max Planck Society, and the Higher
Education Funding Council for England.  The SDSS Web site \hbox{is
{\tt http://www.sdss.org/}.}


\begin{thebibliography}{}

\bibitem[Abazajian et al.(2005)]{2005AJ....129.1755A} Abazajian, K., et 
al.\ 2005, \aj, 129, 1755 

\bibitem[Akritas \& Bershady(1996)]{1996ApJ...470..706A} Akritas, M.~G.~\& 
Bershady, M.~A.\ 1996, \apj, 470, 706 

\bibitem[Alexander (1997)]{A1987} Alexander, T. 1997, in Astronomical
Time Series, ed. D. Maoz, A. Sternberg, \& E. M. Leibowitz (Dordrecht:
Kluwer), 163

\bibitem[Baldwin(1977)]{1977ApJ...214..679B} Baldwin, J.~A.\ 1977, \apj, 
214, 679 

\bibitem[Baskin \& Laor(2005)]{2005MNRAS.356.1029B} Baskin, A., \& Laor, 
A.\ 2005, \mnras, 356, 1029 

\bibitem[Bentz et al.(2006)]{2006ApJ...644..133B} Bentz, M.~C., Peterson, 
B.~M., Pogge, R.~W., Vestergaard, M., \& Onken, C.~A.\ 2006, \apj, 644, 133 

\bibitem[Barger et al.(2005)]{2005AJ....129..578B} Barger, A.~J., Cowie, 
L.~L., Mushotzky, R.~F., Yang, Y., Wang, W.-H., Steffen, A.~T., \& Capak, 
P.\ 2005, \aj, 129, 578 

\bibitem[Borgeest \& Schramm(1994)]{1994A&A...284..764B} Borgeest, U., \& 
Schramm, K.-J.\ 1994, \aap, 284, 764 

\bibitem[Cardelli et al.(1989)]{1989ApJ...345..245C} Cardelli, J.~A., 
Clayton, G.~C., \& Mathis, J.~S.\ 1989, \apj, 345, 245 

\bibitem[Cid Fernandes et al.(2000)]{2000ApJ...544..123C} Cid Fernandes, 
R., Sodr{\'e}, L., Jr., \& Vieira da Silva, L., Jr.\ 2000, \apj, 544, 123 

\bibitem[Clavel et al.(1991)]{1991ApJ...366...64C} Clavel, J., et al.\ 
1991, \apj, 366, 64 

\bibitem[Collier et al.(1998)]{1998ApJ...500..162C} Collier, S.~J., et al.\ 
1998, \apj, 500, 162 

\bibitem[Edelson et al.(2002)]{2002ApJ...568..610E} Edelson, R., Turner, 
T.~J., Pounds, K., Vaughan, S., Markowitz, A., Marshall, H., Dobbie, P., \& 
Warwick, R.\ 2002, \apj, 568, 610 

\bibitem[Gaskell(1994)]{1994ASPC...69..111G} Gaskell, C.~M.\ 1994, in
Reverberation Mapping of the Broad-Line Region in AGNs, ed. P. Gondhalekar,
K. Horne, \& B. M. Peterson (San Francisco: ASP), 111 

\bibitem[Giveon et al.(1999)]{1999MNRAS.306..637G} Giveon, U., Maoz, D., 
Kaspi, S., Netzer, H., \& Smith, P.~S.\ 1999, \mnras, 306, 637 

\bibitem[Hill et al.(1998)]{1998SPIE.3355..375H} Hill, G.~J., Nicklas, 
H.~E., MacQueen, P.~J., Tejada, C., Cobos Duenas, F.~J., \& Mitsch, W.\ 
1998, \procspie, 3355, 375 

\bibitem[Jester et al.(2005)]{2005AJ....130..873J} Jester, S., et al.\ 
2005, \aj, 130, 873 

\bibitem[Jiang et al.(2006)]{2006astro.ph.11453J} Jiang, L., Fan, X., 
Ivezic, Z., Richards, G.~T., Schneider, D.~P., Strauss, M.~A., \& Kelly, 
B.~C.\ 2006, ApJ, in press, astro-ph/0611453 

\bibitem[Kaspi et al.(1995)]{K1995} Kaspi, S., Ibbetson P.~A., Mashal, E., \&
Brosch, N. 1995, Wise Obs. Tech. Rep. 6

\bibitem[Kaspi et al.(1996)]{1996ApJ...470..336K} Kaspi, S., et al.\ 1996a, 
\apj, 470, 336 

\bibitem[Kaspi et al.(1996)]{1996ApJ...471L..75K} Kaspi, S., Smith, P.~S., 
Maoz, D., Netzer, H., \& Jannuzi, B.~T.\ 1996b, \apjl, 471, L75 

\bibitem[Kaspi et al.(2000)]{2000ApJ...533..631K} Kaspi, S., Smith, P.~S., 
Netzer, H., Maoz, D., Jannuzi, B.~T., \& Giveon, U.\ 2000, \apj, 533, 631 

\bibitem[Kaspi et al.(2005)]{2005ApJ...629...61K} Kaspi, S., Maoz, D., 
Netzer, H., Peterson, B.~M., Vestergaard, M., \& Jannuzi, B.~T.\ 2005, 
\apj, 629, 61 

\bibitem[Kollmeier et al.(2005)]{2005astro.ph..8657K} Kollmeier, J.~A., et 
al.\ 2006, \apj, 648, 128      

\bibitem[Korista et al.(1995)]{1995ApJS...97..285K} Korista, K.~T., et al.\ 
1995, \apjs, 97, 285 

\bibitem[Laor \& Brandt (2002)]{1002laor} Laor, A., \& 
Brandt, W.~N.\ 2002, \apj, 569, 641

\bibitem[Lawrence \& Papadakis(1993)]{1993ApJ...414L..85L} Lawrence, A., \& 
Papadakis, I.\ 1993, \apjl, 414, L85 

\bibitem[Marconi et al.(2004)]{2004MNRAS.351..169M} Marconi, A., Risaliti, 
G., Gilli, R., Hunt, L.~K., Maiolino, R., \& Salvati, M.\ 2004, \mnras, 
351, 169 

\bibitem[Markowitz et al.(2003)]{2003ApJ...593...96M} Markowitz, A., et 
al.\ 2003, \apj, 593, 96 

\bibitem[Maoz \& Netzer (1989)]{MN89} Maoz, D., \& Netzer, H. 1989,
MNRAS, 236, 21

\bibitem[Maoz et al.(1993)]{1993ApJ...409...28M} Maoz, D., et al.\ 1993, 
\apj, 409, 28 

\bibitem[Maoz et al.(1994)]{1994ApJ...421...34M} Maoz, D., Smith, P.~S., 
Jannuzi, B.~T., Kaspi, S., \& Netzer, H.\ 1994, \apj, 421, 34 

\bibitem[Maoz et al.(2002)]{2002AJ....124.1988M} Maoz, D., Markowitz, A., 
Edelson, R., \& Nandra, K.\ 2002, \aj, 124, 1988 

\bibitem[McLure \& Dunlop(2004)]{2004MNRAS.352.1390M} McLure, R.~J., \& 
Dunlop, J.~S.\ 2004, \mnras, 352, 1390 

\bibitem[McLure \& Jarvis(2002)]{2002MNRAS.337..109M} McLure, R.~J.~\& 
Jarvis, M.~J.\ 2002, \mnras, 337, 109 

\bibitem[Metzroth et al.(2006)]{2006astro.ph..5038M} Metzroth, K.~G., 
Onken, C.~A., \& Peterson, B.~M.\ 2006, \apj, 647, 901 

\bibitem[Morgan et al.(2003)]{2003AJ....126..696M} Morgan, N.~D., Gregg, 
M.~D., Wisotzki, L., Becker, R., Maza, J., Schechter, P.~L., \& White, 
R.~L.\ 2003, \aj, 126, 696 

\bibitem[Netzer(2003)]{2003ApJ...583L...5N} Netzer, H.\ 2003, \apjl, 583, L5 

\bibitem[Netzer et al.(1996)]{1996MNRAS.279..429N} Netzer, H., et al.\ 
1996, \mnras, 279, 429 

\bibitem[Netzer \& Peterson(1997)]{1997ats..proc...85N} Netzer, H.~\& 
Peterson, B.~M.\ 1997, in Astronomical Time Series, ed. D. Maoz, A. Sternberg
and E. Leibowitz (Dordrecht: Kluwer Academic Publishers), 85

\bibitem[O'Brien et al.(1998)]{1998ApJ...509..163O} O'Brien, P.~T., et al.\ 
1998, \apj, 509, 163 

\bibitem[O'Neill et al.(2005)]{2005MNRAS.358.1405O} O'Neill, P.~M., Nandra, 
K., Papadakis, I.~E., \& Turner, T.~J.\ 2005, \mnras, 358, 1405 

\bibitem[Peterson(1993)]{1993PASP..105..247P} Peterson, B.~M.\ 1993, \pasp, 
105, 247 

\bibitem[Peterson(2006)]{p2006} Peterson, B.~M.\ 2006, in Physics
of Active Galactic Nuclei at All Scales, ed. D. Alloin, R. Johnson,
P. Lira (Berlin: Springer-Verlag), in press

\bibitem[Peterson et al.(1998a)]{1998ApJ...501...82P} Peterson, B.~M., 
Wanders, I., Bertram, R., Hunley, J.~F., Pogge, R.~W., \& Wagner, R.~M.\ 
1998a, \apj, 501, 82 

\bibitem[Peterson et al.(1998b)]{1998PASP..110..660P} Peterson, B.~M., 
Wanders, I., Horne, K., Collier, S., Alexander, T., Kaspi, S., \& Maoz, D.\ 
1998b, \pasp, 110, 660 

\bibitem[Peterson et al.(2004)]{P2004} Peterson, B.~M., et
al.\ 2004, ApJ, 613, 682

\bibitem[Peterson et al.(2005)]{2005astro.ph..6665P} Peterson, B.~M., et
al.\ 2005, ApJ, 632, 799. Erratum: 2006, ApJ, 641, 638

\bibitem[Press et al. (1992)]{press1992} Press, W. H., Teukolsky,
S. A., Vetterling, W. T, \& Flannery, B. P. 1992, Numerical Recipes
in FORTRAN (Second ed.; Cambridge: Cambridge Univ. press)

\bibitem[Ramsey et al.(1998)]{1998SPIE.3352...34R} Ramsey, L.~W., et al.\ 
1998, \procspie, 3352, 34 

\bibitem[Reichert et al.(1994)]{1994ApJ...425..582R} Reichert, G.~A., et 
al.\ 1994, \apj, 425, 582 

\bibitem[Reimers et al.(2005)]{2005A&A...436..465R} Reimers, D., Hagen, 
H.-J., Schramm, J., Kriss, G.~A., \& Shull, J.~M.\ 2005, \aap, 436, 465

\bibitem[Riess et al.(2004)]{2004ApJ...607..665R} Riess, A.~G., et al.\ 
2004, \apj, 607, 665 
 
\bibitem[Rodriguez-Pascual et al.(1997)]{1997ApJS..110....9R} 
Rodriguez-Pascual, P.~M., et al.\ 1997, \apjs, 110, 9 

\bibitem[Schlegel et al.(1998)]{1998ApJ...500..525S} Schlegel, D.~J., 
Finkbeiner, D.~P., \& Davis, M.\ 1998, \apj, 500, 525 

\bibitem[Schmidt \& Green(1983)]{1983ApJ...269..352S} Schmidt, M., \& 
Green, R.~F.\ 1983, \apj, 269, 352 
 
\bibitem[Schneider et al.(2005)]{2005AJ....130..367S} Schneider, D.~P., et 
al.\ 2005, \aj, 130, 367 

\bibitem[Shaviv(1998)]{1998ApJ...494L.193S} Shaviv, N.~J.\ 1998, \apjl, 
494, L193

\bibitem[Shemmer et al.(2003)]{2003MNRAS.343.1341S} Shemmer, O., Uttley, 
P., Netzer, H., \& McHardy, I.~M.\ 2003, \mnras, 343, 1341 

\bibitem[Shemmer et al.(2004)]{2004ApJ...614..547S} Shemmer, O., Netzer, 
H., Maiolino, R., Oliva, E., Croom, S., Corbett, E., \& di Fabrizio, L.\ 
2004, \apj, 614, 547 

\bibitem[Spergel et al.(2003)]{2003ApJS..148..175S} Spergel, D.~N., et al.\ 
2003, \apjs, 148, 175 

\bibitem[Steffen et al.(2006)]{2006AJ....131.2826S} Steffen, A.~T., 
Strateva, I., Brandt, W.~N., Alexander, D.~M., Koekemoer, A.~M., Lehmer, 
B.~D., Schneider, D.~P., \& Vignali, C.\ 2006, \aj, 131, 2826 

\bibitem[Strateva et al.(2005)]{2005AJ....130..387S} Strateva, I.~V., Brandt, 
W.~N., Schneider, D.~P., Vanden Berk, D.~G., \& Vignali, C.\ 2005, \aj, 1

\bibitem[Tremaine et al.(2002)]{2002ApJ...574..740T} Tremaine, S., et al.\ 
2002, \apj, 574, 740 

\bibitem[Trevese et al.(2004)]{2004astro.ph..8075T} Trevese,
D., Stirpe, G., Vagnetti, F., Zitelli, V., \& Paris, D.\ 2004,
proceedings of "AGN Variability from X-rays to Radio Waves",
Eds. C. M. Gaskell, I. M. McHardy, B. M. Peterson, and S. G. Sergeev,
in press, astro-ph/0408075

\bibitem[Ulrich et al.(1993)]{1993ApJ...411..125U} Ulrich, M.-H., 
Courvoisier, T.~J.-L., \& Wamsteker, W.\ 1993, \apj, 411, 125 

\bibitem[Uttley et al.(2002)]{2002MNRAS.332..231U} Uttley, P., McHardy, 
I.~M., \& Papadakis, I.~E.\ 2002, \mnras, 332, 231 

\bibitem[Vanden Berk et al.(2004)]{2004ApJ...601..692V} Vanden Berk, D.~E., 
et al.\ 2004, \apj, 601, 692 

\bibitem[Veron-Cetty \& Veron(1993)]{1993cqan.book.....V} Veron-Cetty, 
M.-P., \& Veron, P.\ 1993, ESO Scientific Report, Garching: European 
Southern Observatory (ESO), 6th ed.

\bibitem[Vestergaard (2002)]{2002ApJ...571..733V} Vestergaard, M.\ 2002, 
\apj, 571, 733 

\bibitem[Vestergaard \& Peterson(2006)]{2006ApJ...641..689V} Vestergaard, 
M., \& Peterson, B.~M.\ 2006, \apj, 641, 689 

\bibitem[Wanders et al.(1997)]{1997ApJS..113...69W} Wanders, I., et al.\ 
1997, \apjs, 113, 69 

\bibitem[Welsh et al.(2000)]{2000AAS...197.3913W} Welsh, W., et al.\ 2000, 
\baas , 32, 1458 

\bibitem [White \& Peterson (1994)]{WP94} White, R. J., \& Peterson,
B. M. 1994, PASP, 106, 879

\bibitem[Wills et al.(1993)]{1993ApJ...410..534W} Wills, B.~J., Netzer, H., 
Brotherton, M.~S., Han, M., Wills, D., Baldwin, J.~A., Ferland, G.~J., \& 
Browne, I.~W.~A.\ 1993, \apj, 410, 534 

\bibitem[Woo \& Urry(2002)]{2002ApJ...581L...5W} Woo, J.-H., \& Urry, 
C.~M.\ 2002, \apjl, 581, L5 

\bibitem[Wu et al.(2004)]{2004A&A...424..793W} Wu, X.-B., Wang, R., Kong, 
M.~Z., Liu, F.~K., \& Han, J.~L.\ 2004, \aap, 424, 793 

\bibitem[York et al.(2000)]{2000AJ....120.1579Y} York, D.~G., et al.\ 2000, 
\aj, 120, 1579 

\bibitem[Yu et al.(2005)]{2005ApJ...634..901Y} Yu, Q., Lu, Y., \& 
Kauffmann, G.\ 2005, \apj, 634, 901 

\end{thebibliography}
\end{document}